\documentclass{easychair}

\usepackage{doc}
\usepackage{listings}
\usepackage{subfig}
\usepackage{graphicx}
\usepackage{amsmath}
\usepackage{framed}
\usepackage{xcolor}
\usepackage{enumitem}

\usepackage{wrapfig}

\usepackage{amsfonts}
\usepackage{makecell}
\usepackage{multirow}
\usepackage{url}
\usepackage{hyperref}
\usepackage{array}

\usepackage{newfloat}

\DeclareFloatingEnvironment[fileext=lop]{listing}

\makeatletter
\newcommand{\printfnsymbol}[1]{%
  \textsuperscript{\@fnsymbol{#1}}%
}
\makeatother

\newif\ifcomments
\commentstrue

\usepackage{lineno}
\usepackage[ruled,linesnumbered,noend]{algorithm2e}

\title{\textit{HTBQueue}: A Hierarchical Token Bucket Implementation for the OMNeT++/INET Framework}

\author{
    Marcin Bosk\inst{1}\textsuperscript{,$\ast$}
\and
    Marija Gaji\'c\inst{2}\textsuperscript{,$\ast$}
\and
   Susanna Schwarzmann \inst{3}
\and   
    Stanislav Lange\inst{2}
\and   
    Thomas Zinner\inst{2} 
}

\institute{
  Technical University of Munich, Department of Informatics, Chair of Connected Mobility\\
\and
   NTNU - Norwegian University of Science and Technology, Department of Information Security and Communication Technology\\
\and
    Technische Universität Berlin, Chair of Internet Network Architectures\\
 }

\authorrunning{Bosk, Gaji\'c, Schwarzmann, Lange, and Zinner}

\titlerunning{\textit{HTBQueue} for OMNeT++/INET}

\begin{document}

\maketitle
\begingroup\renewcommand\thefootnote{*}
\footnotetext{Marcin Bosk and Marija Gaji\'c contributed equally to this paper.}
\begingroup\renewcommand\thefootnote{\arabic{footnote}}
\begin{abstract}
The hierarchical token bucket~(HTB) algorithm allows to specify per-flow bitrate guarantees and enables excess bandwidth sharing between flows of the same class. Additionally, it provides capabilities to prioritize the traffic of specific flows,  potentially considering their delay demands.  
HTB hence constitutes a powerful mechanism to enforce QoS requirements hierarchically and on a fine granular per-flow level, making it an appropriate choice in numerous use-cases. 
In this paper, we present \mbox{\textit{HTBQueue}}, our implementation of a compound module for HTB support in the discrete event simulator OMNeT++.
We validate \mbox{\textit{HTBQueue}}'s functionality in terms of rate conformance and fair bandwidth sharing behavior between competing flows. We furthermore demonstrate its support for flow prioritization.  
\end{abstract}


\section{Introduction}
\label{sec:intro}

The Hierarchical Token Bucket (HTB) algorithm is one of the most widely used mechanisms for rate limiting. Implemented in the Linux traffic control tool \texttt{tc}, it is used for network control in testbed setups for research experiments~\cite{chun2003planetlab,li2016emustack,yan2015vt} or in general to enforce per-flow Quality of Service (QoS) policies~\cite{valenzuela_htbQos,vestin2015qos,ren2017end}. HTB allows to classify various types of traffic into different queues, according to properties such as the service type or IP address. The different queues into which traffic is classified can be configured with a priority level and a rate limit, allowing to schedule the packets and shape the traffic accordingly. HTB relies on two token buckets for controlling the bandwidth usage of a link.  
The tokens are generated with the desired per-flow rate and packets can only be dequeued if enough tokens are available in the bucket. 
The unique feature of HTB is the bandwidth sharing mechanism. 
A flow which does not fully exploit its \textit{assured rate} consequently does not consume all of its available tokens. 
These excess tokens (representing the excess bandwidth) can be borrowed by other flows sharing the same parent in the HTB structure. 
The amount of bandwidth a flow can borrow is limited by its \textit{ceiling rate}. 

Although the concept of HTB and its features of setting assured and ceiling rate are relatively old, the same mechanisms are still proposed in modern networking architectures. 
The 3GPP~\cite{3gppts38300} standard for the Next Generation Radio Access Network (NG-RAN) in 5G defines a guaranteed bitrate~(GBR) as well as a maximum bitrate~(MBR) for each flow. 
While the C++ based discrete event simulator OMNeT++~\cite{omnetOverview} 
implements a generic module for token buckets\footnote{\url{https://inet.omnetpp.org/docs/tutorials/queueing/doc/TokenBucket.html}}, a module providing the full capabilities of HTB, including per-flow assured and ceiling rate, is still missing. 
In this paper, we propose a
ready-to-use OMNeT++ module\footnote{Implementation available at: \url{https://github.com/fg-inet/omnet_htb}} for the INET Framework~\cite{inetFrame} that supports two-level bitrate guarantees. Our compound module, called \textit{HTBQueue} implements HTB's classful queueing approach based on the Linux HTB implementation~\cite{linuxHTB}. 
It allows to set priorities and rate limits for an arbitrary number of flows.\footnote{Due to the simulative nature our implementation does not have scalability issues, while Linux HTB does.}
With the presented tool, we can define a maximum rate for the root and include several intermediate levels that define how the overall capacity is allocated among the different classes and leaf nodes and to which extent leaf nodes can borrow from each other. 

The remainder of the paper is structured as follows.
Section~\ref{sec:relatedwork} presents related work. Section~\ref{sec:htb_overview} gives a brief and general overview on the working principles of HTB, while Section~\ref{sec:implementation} details on our specific implementation. We validate the functionality of our compound module in Section~\ref{sec:eval}. 
Finally, Section~\ref{sec:conclusion} concludes the paper.

\section{Related Work}
\label{sec:relatedwork}

According to \cite{linuxHTB}, the main concept and idea of the HTB originate from~\cite{floyd_link_sharing} where a hierarchical model for link sharing and a resource management model were presented. In~\cite{valenzuela_htbQos}, the HTB was used to enhance the QoS guaranteed in wireless local networks. It was shown that very low standard deviation and sustainable throughput in such networks are achievable via application of the HTB traffic control in the MAC layer. In addition, this paper shows that HTB introduces hierarchical classes. Such classes are highly desired by the operators due to the fact that they allow for more efficient and scalable network management compared with the per-flow traffic control. High rate conformance of the HTB was further tested and confirmed  in~\cite{saed_htb} and~\cite{ivancic2005}. In \cite{balan2009}, the authors show the potential of the wide range of the actual Linux-based HTB implementations (including the CLI method, Linux Network Service, text configuration files, and the Web interface) used for satisfying complex QoS requirements. 
The Linux implementation on which we based our compound OMNeT++ module was presented in~\cite{linuxHTB} and~\cite{htb_howto}.
In our recent work~\cite{bosk2021using}, one of the use cases of the proposed HTB OMNeT++ module was presented. We investigated the potentials of using QoS Flows and slicing to achieve high customer QoE and better utilization of the available resources. Our implementation of the HTB in OMNeT++ was used to emulate and evaluate slice-like traffic isolation.

Another area where traffic shaping plays a key role in meeting performance guarantees is time-sensitive networking~(TSN). In this context, various shaping mechanisms such as the credit-based shaper are standardized and have been implemented as OMNeT++ modules within the NeSTiNg~\cite{falk2019nesting} framework. However, in contrast to the HTB presented in this work, TSN-related shapers usually do not provide means of two-level shaping and borrowing via guaranteed and maximum bitrate settings.



\section{HTB Overview}
\label{sec:htb_overview}

Traffic control is an essential part of today's network management, supporting a myriad of coexisting heterogeneous applications. According to the 3GPP standards for the Next Generation Radio Access Networks (NG-RAN) in 5G \cite{3gppts38300}, each flow should have its \textit{guaranteed (GBR)} and \textit{maximum bitrate (MBR)}. HTB is a traffic shaper and policer which allows such two-level bitrate guarantees and is therefore very useful to mobile network operators. In the HTB terminology, the GBR and MBR precisely translate to the \textit{assured bitrate} and \textit{ceiling bitrate}. 

HTB is a type of classful \textit{token bucket} algorithm. The rate of all the \textit{classes} in the HTB hierarchy is controlled with two nested token buckets governed respectively by \textit{tokens} and \textit{ctokens}. The \textit{tokens} are for sending at the assured rate, and the \textit{ctokens} are for sending at the ceiling rate.  By a \textit{class}, we refer to a type of node in the HTB tree structure. 
 %

\begin{wrapfigure}{L}{0.50\textwidth}
\centering
\includegraphics[width=0.50\textwidth]{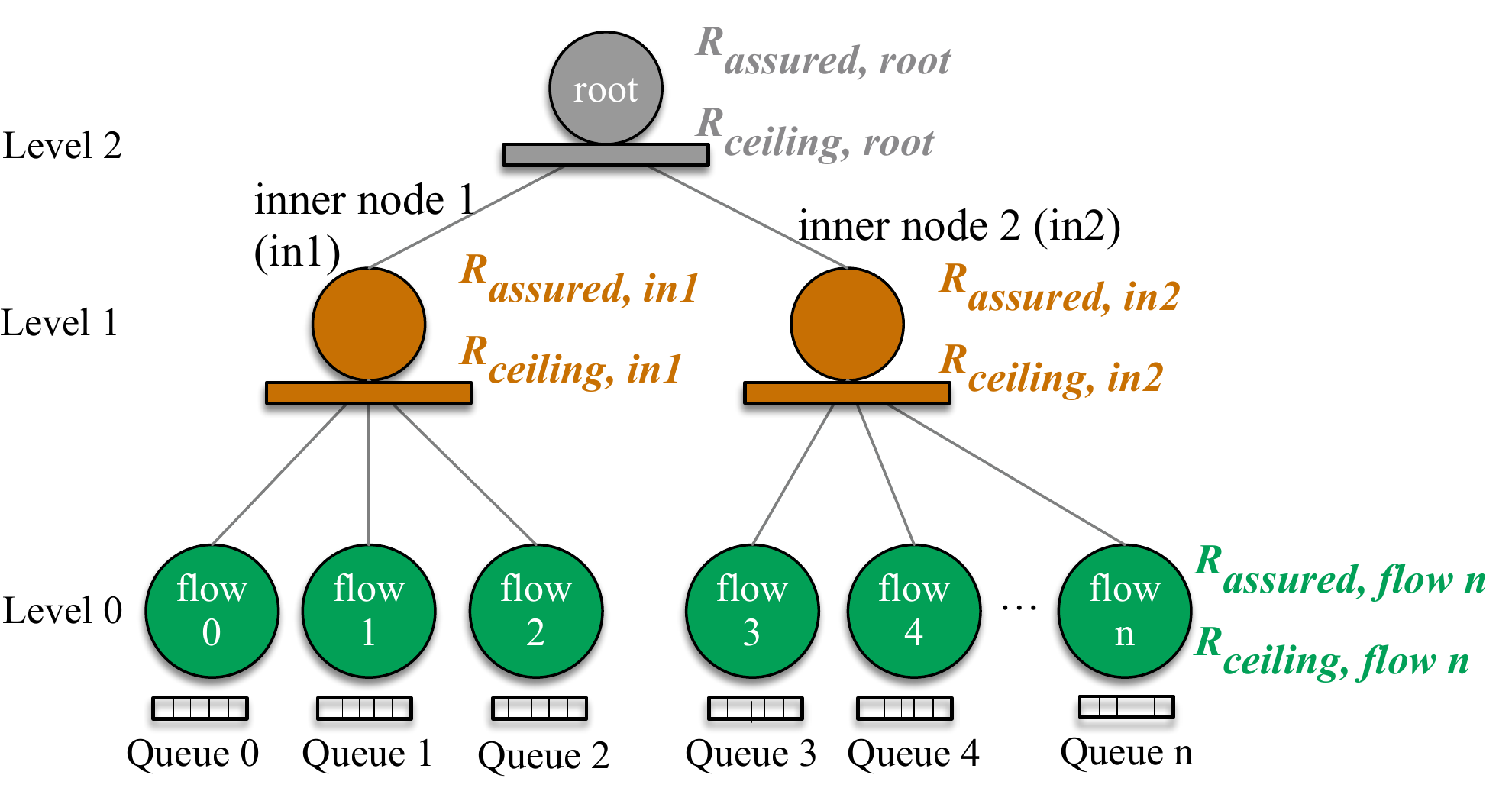}
\caption{\label{fig:htb_tree} Exemplary HTB tree structure.}
\end{wrapfigure}
\noindent
The key structure for the HTB hierarchy is the HTB tree. It consists of three types of classes (or tree nodes): root, intermediate/inner nodes, and leaves. 
The root node has no parent node, while the leaves have no child nodes. Inner nodes are the intermediate nodes between the root and the leaves. An example of such a tree structure with 3 levels, $L_{max}=2$, is shown in Figure~\ref{fig:htb_tree}.  The leaves are on the lowest level $L=0$, and the root is on the highest level $L=2$. In the figure, we have two inner nodes constituting the middle level $L=1$. One leaf can be thought of as one flow. In general, only the leaves have queues. Each leaf has an associated priority level - the lower the value, the higher the priority. 

All three HTB classes have the following main parameters: the level to which they belong, their assured rate $R_{assured}$, their ceiling rate $R_{ceiling}$ which represents the maximum achievable rate, and a quantum $Q_{class}$ that defines the maximum size of data that can be dequeued in one round. A prerequisite for the rates is that the sum of the assured rates of one node's children has to be less than or equal to the assured rate of that node. The root's assured and ceiling rate are usually set equal to the link bandwidth. 
Depending on a node's current rate $R_{current}$, at every instance of time, the state of the node is determined with the following three values: 

\begin{itemize}[noitemsep]
    \setlength\itemsep{0em}
    \item 0 - \textit{can send}: $R_{current}\leq R_{assured},$
    \item 1 - \textit{may borrow}: $R_{assured} < R_{current} < R_{ceiling} $, and
    \item 2 - \textit{can't send}: $R_{current}\geq R_{ceiling}$.
\end{itemize}

In addition, each class on level $L>0$ keeps a list $D_{class}$ of its descendants that are in \textit{may borrow} mode, including both leaves and intermediate nodes (cf. Figure~\ref{fig:htb_tree}). Descendants in this list can, the same as in Linux HTB, have different priority levels $P \in \mathbb{Z}^+, P \leq 7$. 

The fundamental working principle of the HTB is to first satisfy the assured rate of all active classes on all priority levels, and then share the parent's excess bandwidth fairly among all the child nodes that are in \textit{may borrow} mode. Among the nodes in this list, those with the highest priority are served first. The portion of the excess bandwidth that each descendant class gets from its parent is \textit{the borrowed bandwidth} - $B_{class}$. The same principle is then recursively applied for lower priority levels. This link sharing principle is more precisely described with the following equations\footnote{Based on equations from: \url{http://luxik.cdi.cz/~devik/qos/htb/manual/theory.htm}, accessed on 09.06.21}: 


\begin{equation}
\label{eq:link_share1}
\small
R_{class}= \min(R_{ceiling}, R_{assured}+B_{class})
\end{equation}
\begin{equation}
\label{eq:link_share2}
    B_{class}=
\begin{cases}
    \frac{Q_{class} \cdot R_{parent}}{\sum_{i\in\{x\in D_{parent}\vert P_x=P_{class}\}}Q_i}, & \text{if } \min(\{P_{x}\vert x\in D_{parent}\}) \geq P_{class} \\
    0,              & \text{otherwise}
\end{cases}
\end{equation}
where $R_{parent}$ is the current rate utilized by the parent, $Q_{P}$ are quantums of the descendant nodes in $D_{parent}$, which are on the same priority level as the class of interest ($P=P_{class}$). The HTB based scheduler applies these equations always starting from the active classes at the lowest tree level ($L=0$).


From Equations~\ref{eq:link_share1} and~\ref{eq:link_share2}, we can see that formulas are recursive 
($R_{parent}$ is determined via Equation~\ref{eq:link_share1}), as well as that prioritized applications (i.e., the ones with lower $P$ values) are being served first. In the cases when there are several descendant nodes borrowing from a common parent and having the same priority level, Deficit Round Robin~(DRR) is applied to prevent potential starvation of the classes~\cite{drr}. The amount of bytes sent per node and DRR round is controlled by the quantum. 

\section{Implementation} 
\label{sec:implementation}
We base our implementation on existing queueing compound modules from the OMNeT++ INET framework. 
More specifically, we extend the \textit{PriorityQueue} to build our compound module. 
\textit{HTBQueue} consists of two main modules: a classifier (\textit{HTBClassifier}) and a scheduler (\textit{HTBScheduler}). Packets enqueued by the HTB are stored within multiple queuing modules.
Any queue from the INET framework can be used with the default being the generic \textit{PacketQueue}.
As the implementation requires link-layer specific knowledge and a slight adjustment to the utilized interface module (see Section \ref{subsec:schedule}), currently, the \textit{HTBQueue} module can be used only with the Point-to-Point (PPP) interface available within the INET framework. 
We use OMNeT++ version 5.5.1\footnote{\url{https://github.com/omnetpp/omnetpp/releases/tag/omnetpp-5.5.1}, last accessed on 15th of June 2021} and INET Framework version 4.2.0\footnote{\url{https://github.com/inet-framework/inet/releases/tag/v4.2.0}, last accessed on 15th of June 2021}.

\subsection{\textit{HTBQueue} Compound Module Functional Overview}
\label{sec:model_overview}
Figure~\ref{fig:omnet_htb} presents the functional flow of the \textit{HTBQueue} module operation from packet arrival until packet dequeue.
Upon arrival \raisebox{.5pt}{\textcircled{\raisebox{-.9pt} {1}}}, the classifier classifies all incoming packets according to its filters.
These packets are placed \raisebox{.5pt}{\textcircled{\raisebox{-.9pt} {2}}} into a respective queue that directly corresponds to a leaf in the HTB class structure. The classifier also informs \raisebox{.5pt}{\textcircled{\raisebox{-.9pt} {3}}} the scheduler into which queue a packet is placed. 
It allows the \textit{HTBScheduler} to activate the respective classes. This way the scheduler knows that there are packets present in the queue of the respective leaf class and it can include these classes in the dequeue operation. 
The packets stay in their queues until the PPP interface sends \raisebox{.5pt}{\textcircled{\raisebox{-.9pt} {4}}} a \textit{ready-to-send} signal. 
Upon reception of that signal \raisebox{.5pt}{\textcircled{\raisebox{-.9pt} {5}}}, the scheduler determines the index of the next leaf (and in turn the index of the packet queue) to dequeue. 
The queue index is \raisebox{.5pt}{\textcircled{\raisebox{-.9pt} {6}}} determined based on the information that the scheduler has about the queues, their rate boundaries, class states, and the DRR resource sharing principle.
Lastly, the packets are popped \raisebox{.5pt}{\textcircled{\raisebox{-.9pt} {7}}} from the chosen queue and sent \raisebox{.5pt}{\textcircled{\raisebox{-.9pt} {8}}} over the PPP interface.    
\label{subsec:htb}
\begin{figure*}
    \center{\includegraphics[width=0.78\textwidth]
        {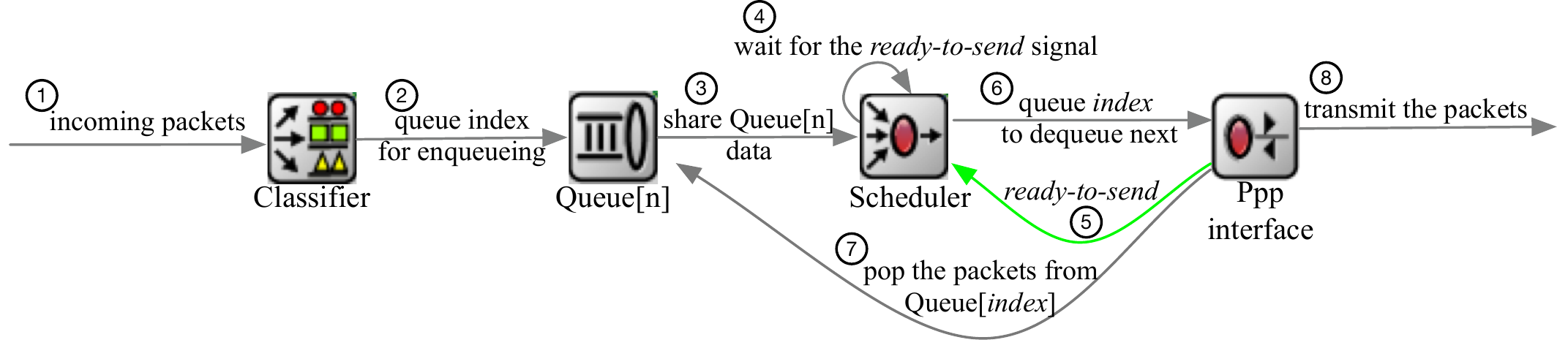}}
        \caption{Structure of the HTBQueue OMNeT++ module.}
        \label{fig:omnet_htb}
\end{figure*}

\subsection{\textit{HTBClassifier} Module}
The \textit{HTBClassifier} module is an adaptation of the already existing \textit{ContentBasedClassifier}.\footnote{For more details, see: \url{https://doc.omnetpp.org/inet/api-current/neddoc/inet.queueing.classifier.ContentBasedClassifier.html}, last accessed on 13th of July 2021}  
The functionality of the \textit{ContentBasedClassifier} is fully preserved and extended to allow for cooperation with the \textit{HTBScheduler} module. 
With the \textit{HTBClassifier}, filters that direct flows into desired leaf classes of the HTB structure can be specified. 
The packets from these flows are subsequently placed in the respective packet queues within the \textit{HTBQueue} module.

The additional functionality compared to the \textit{ContentBasedClassifier} allows the \textit{HTBClassifier} to inform the scheduler into which queue (i.e., leaf class) an incoming packet is placed.
This involves calling the \texttt{htbEnqueue} method of the scheduler which activates the respective leaf class if it is not already active, and thus makes sure that the leaf is considered for the dequeue operation. 


\subsection{\textit{HTBScheduler} Module}
\label{subsec:schedule}

We implement the core functionality of the HTB queueing discipline in the \textit{HTBScheduler} module. 
The module is essentially a port of the Linux HTB source code\footnote{\url{https://github.com/torvalds/linux/blob/master/net/sched/sch_htb.c}, last accessed on 13.07.21}.
In its majority,  the \textit{HTBScheduler} is a translation of the \texttt{C}-based Linux HTB implementation (including all the main functionalities and constructs) into \texttt{C++}/OMNeT++ compatible code.
The \textit{HTBScheduler} does not implement any actual queueing, instead it only keeps track of the states of the packet queues. 
It includes the actual tree structure of the HTB and implements key functions of the HTB responsible for tracking the packet enqueues, keeping the current state in each leaf queue, and selecting a leaf class queue to transmit next.
The dequeue occurs when the PPP interface finished transmission, or is idle and a new packet becomes available for dequeuing.

\begin{figure*}
    \centering
    \includegraphics[width=1.0\textwidth]{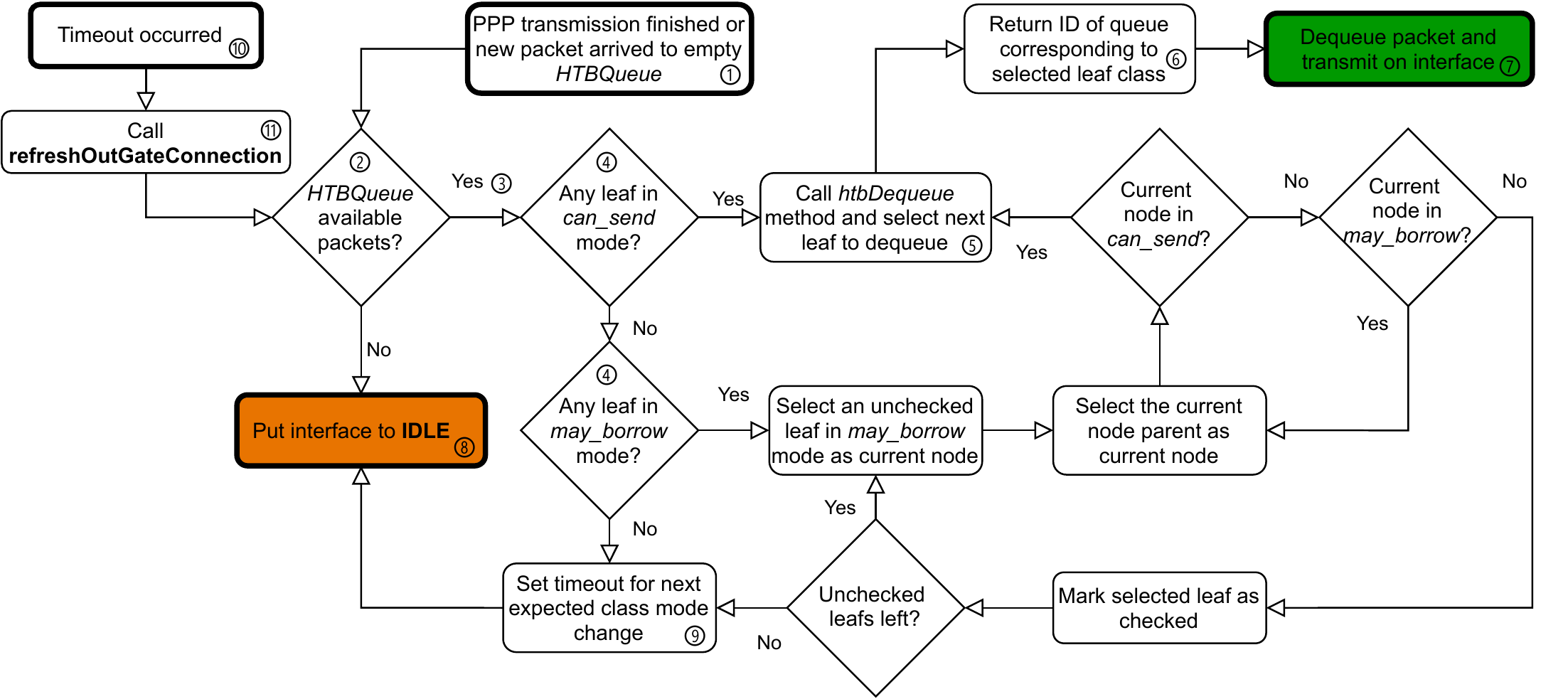}
    \caption{Flow chart of the \textit{HTBScheduler} functionality}
    \label{fig:htbScheduler}
\end{figure*}

The functionality of the \textit{HTBScheduler} is outlined in Figure \ref{fig:htbScheduler}. After a finished transmission~\raisebox{.5pt}{\textcircled{\raisebox{-.9pt} {1}}}, the \textit{PPP} module invokes the \textit{HTBQueue} and checks for packets to dequeue \raisebox{.5pt}{\textcircled{\raisebox{-.9pt} {2}}}. 
If packets are available \raisebox{.5pt}{\textcircled{\raisebox{-.9pt} {3}}}, and any class related to any packet queue with available packets is not in \textit{can't send} mode \raisebox{.5pt}{\textcircled{\raisebox{-.9pt} {4}}}, the \texttt{htbDequeue} method is called~\raisebox{.5pt}{\textcircled{\raisebox{-.9pt} {5}}} and a packet queue index is returned to the interface \raisebox{.5pt}{\textcircled{\raisebox{-.9pt} {6}}}. 
Then, the packet is dequeued and transmitted \raisebox{.5pt}{\textcircled{\raisebox{-.9pt} {7}}}. 
If packets are available and associated classes are in \textit{can't send} mode, the PPP interface is put to idle~\raisebox{.5pt}{\textcircled{\raisebox{-.9pt} {8}}} and a timeout is prepared~\raisebox{.5pt}{\textcircled{\raisebox{-.9pt} {9}}} that informs the \textit{PPP} module at the moment a packet is ready for dequeue. 
The same operations are performed when the \textit{HTBQueue} is empty and a new packet arrives \raisebox{.5pt}{\textcircled{\raisebox{-.9pt} {1}}}. 
The functionality of the \textit{HTBScheduler} also required to make the \texttt{refreshOutGateConnection} method of the \textit{PPP} module \texttt{public}.
The method is called \raisebox{.5pt}{\textcircled{\raisebox{-.9pt} {11}}} from the \textit{HTBScheduler} module in case there is a packet to dequeue, the timeout occurred, and the interface was already idle at that point \raisebox{.5pt}{\textcircled{\raisebox{-.9pt} {10}}}.

The scheduler can be configured using an XML file that defines the HTB class hierarchy (as shown in Figure~\ref{fig:htb_tree}) along with the class-specific settings. 
Each class is represented as a separate element in the XML file, exemplarily shown for a generic class representation in Listing~\ref{list:xml}. 
Detailed instructions and examples for creation of such XML documents can be found in the GitHub repository.

\begin{wraplisting}{L}{0.43\textwidth}
\centering
\includegraphics[width=0.43\textwidth]{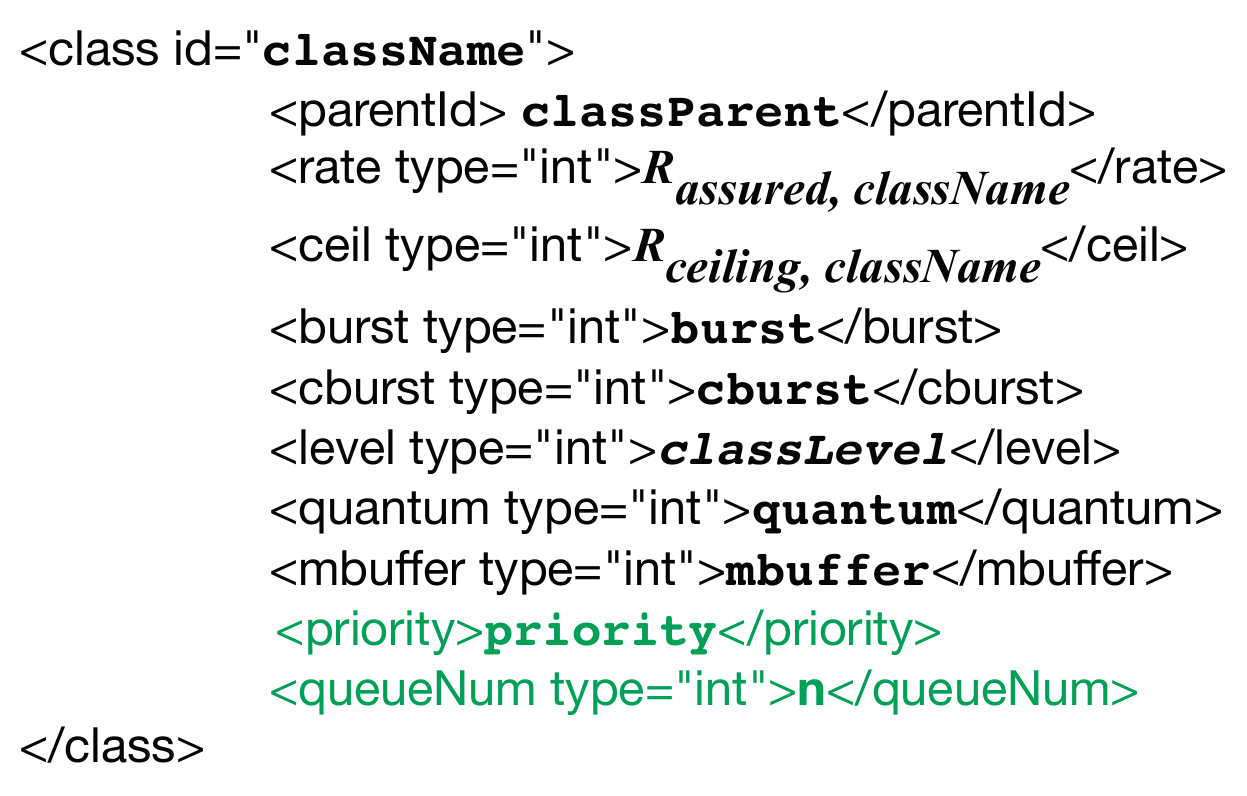}
\caption{XML representation of a class.}
\label{list:xml}
\end{wraplisting}
\noindent
The per class settings include: assured bit-\texttt{rate} ($R_{assured}$), \texttt{ceil}-ing bitrate ($R_{ceiling}$), \texttt{burst} (Bytes that can be burst at MBR), \texttt{cburst} (Bytes that can be burst at link bitrate), \texttt{parentId} (parent in HTB tree hierarchy, \texttt{NULL} for the \texttt{root} class), \texttt{level} (level in the HTB tree structure), \texttt{quantum} (Bytes that can be sent in one DRR round), \texttt{mbuffer} (penalty time for big burst events), \texttt{priority} (only for the leaf classes therefore green in Listing \ref{list:xml}.), and \texttt{queueNum} (corresponding packet queue index, only for the leaf class).
Additionally, each class must have a unique name, the top-most class needs to be called ``root'', and inner as well as leaf classes must have ``inner'' or ``leaf'' in their names respectively.
\section{Validation}\label{sec:evaluation}
\label{sec:eval}

We evaluate the functionality of our HTB implementation via three scenarios: 
Scenario 1 considers five UDP flows without inner classes, Scenario 2 considers five flows with inner classes, and Scenario 3 considers two flows of different priority. 
The simulation setup consists of two hosts, directly connected via a link having 50Mbit/s capacity. The queueing module of the PPP interface on each host is replaced with the \textit{HTBQueue} module. 
The first flow starts at second 0, the subsequent flows start with an offset of 10s each. Each client sends 1500 Byte (physical layer size) packets each 100$\mu$s, corresponding to a bitrate of 120 Mbit/s, and terminates after 100s.  
Each flow is assigned to its own leaf class in the HTB hierarchy. 
The configurations of the HTB and of the rates in the different scenarios are shown in Table \ref{tab:scen_conf}. 
In scenario 2, class Inner 1 is a parent for flows 0, 1, and 2 and Inner 2 is the parent of flows 3 and 4. 
Within scenario 3, flow 0 is prioritized.

\begin{table}[]
\centering
    \caption{Validation scenarios leaf/inner class $R_{assured}$ ($R_a$) and $R_{ceiling}$ ($R_c$) in Mbit/s}
    \label{tab:scen_conf}
    \resizebox{0.85\textwidth}{!}{%
    \begin{tabular}{|c|cc|cc|cc|cc|cc|cc|cc|}
    \hline
     & \multicolumn{2}{c|}{Flow 0} & \multicolumn{2}{c|}{Flow 1} & \multicolumn{2}{c|}{Flow 2} & \multicolumn{2}{c|}{Flow 3} & \multicolumn{2}{c|}{Flow 4} & \multicolumn{2}{c|}{Inner 1} & \multicolumn{2}{c|}{Inner 2} \\
     & $R_a$ & $R_c$ & $R_a$ & $R_c$ & $R_a$ & $R_c$ & $R_a$ & $R_c$ & $R_a$ & $R_c$ & $R_a$ & $R_c$ & $R_a$ & $R_c$ \\ \hline
    Scenario 1 & 3 & 20 & 6 & 25 & 9 & 30 & 12 & 35 & 15 & 40 & \multicolumn{2}{c|}{---} & \multicolumn{2}{c|}{---} \\ \hline
    Scenario 2 & 3 & 20 & 6 & 25 & 9 & 30 & 12 & 35 & 15 & 40 & 20 & 40 & 30 & 40 \\ \hline
    Scenario 3 & 5 & 30 & 5 & 30 & \multicolumn{2}{c|}{---} & \multicolumn{2}{c|}{---} & \multicolumn{2}{c|}{---} & \multicolumn{2}{c|}{---} & \multicolumn{2}{c|}{---} \\ \hline
    \end{tabular}%
}
\end{table}


\begin{figure*}
    \centering
    \subfloat[Scenario 1 - no inner nodes. \label{fig:scenario1}]{
	    \includegraphics[width=0.46\textwidth]{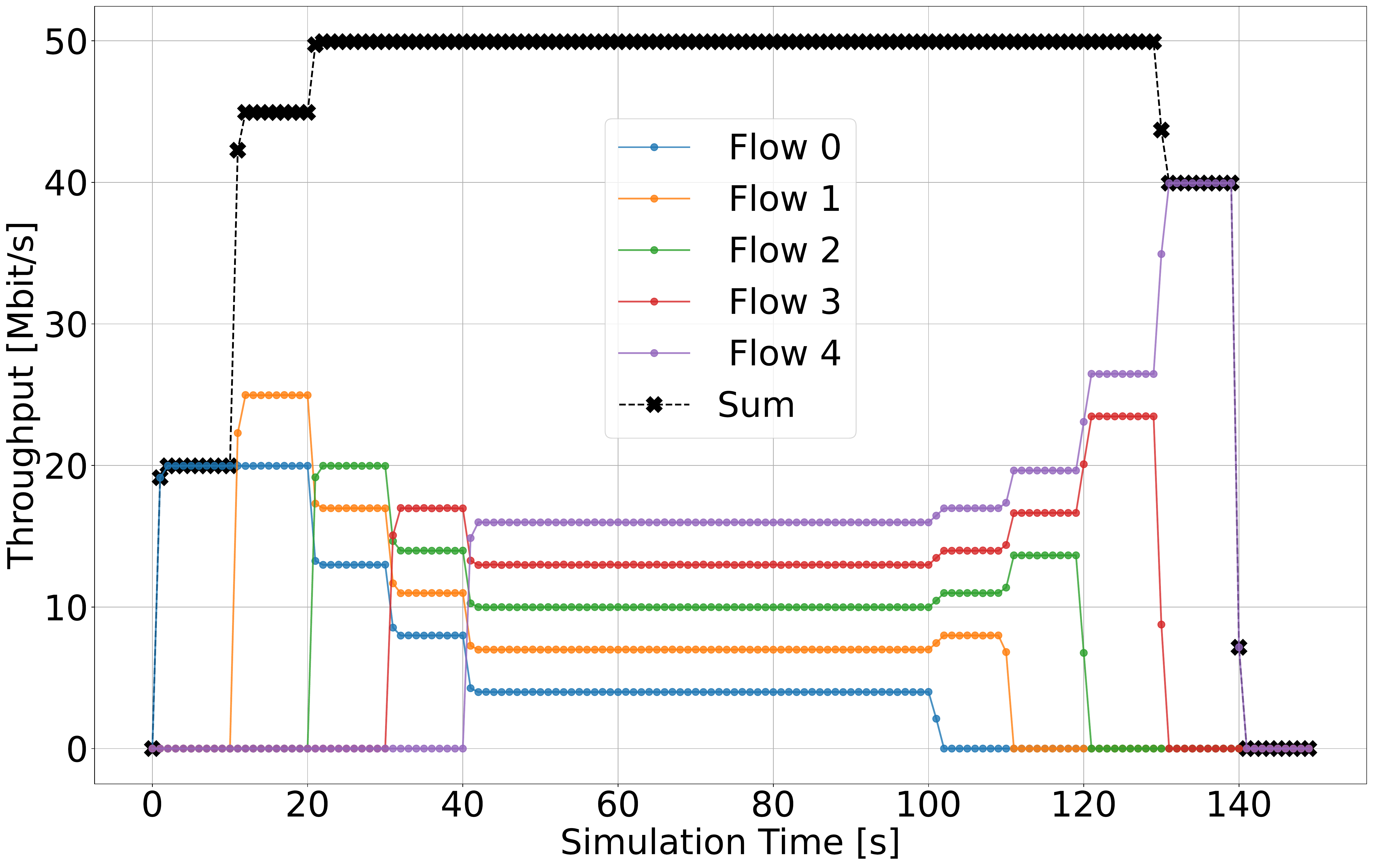}
	}\hfill
	\subfloat[Scenario 2 - two inner nodes. \label{fig:scenario2}]{
	    \includegraphics[width=0.46\textwidth]{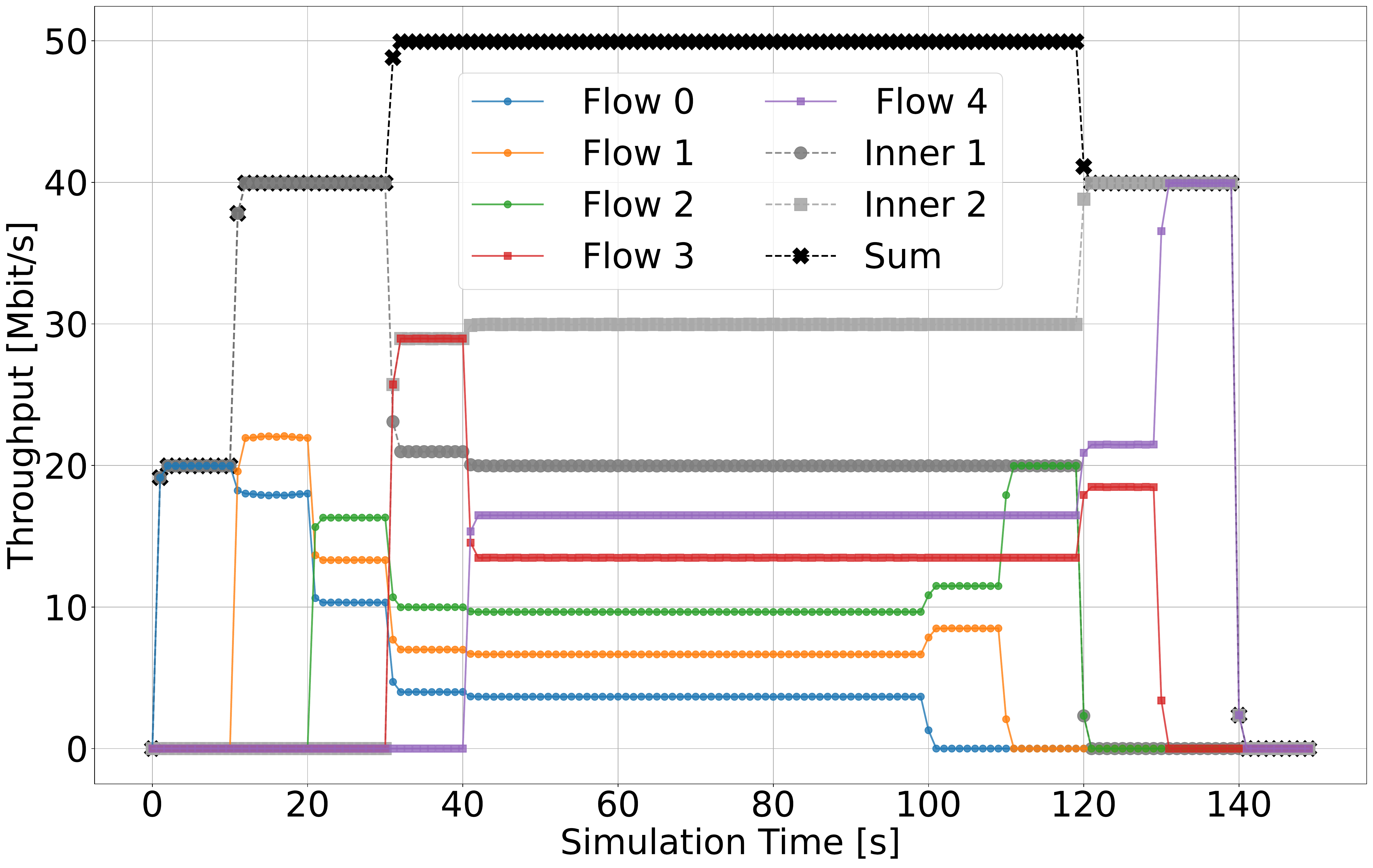}
	}\hfill
    \caption{Scenarios with 5 UDP flows.}
    \label{fig:scenarios12}
\end{figure*}

\begin{figure}
\centering
\begin{minipage}{.46\textwidth}
    \centering
    \includegraphics[width=1\textwidth]{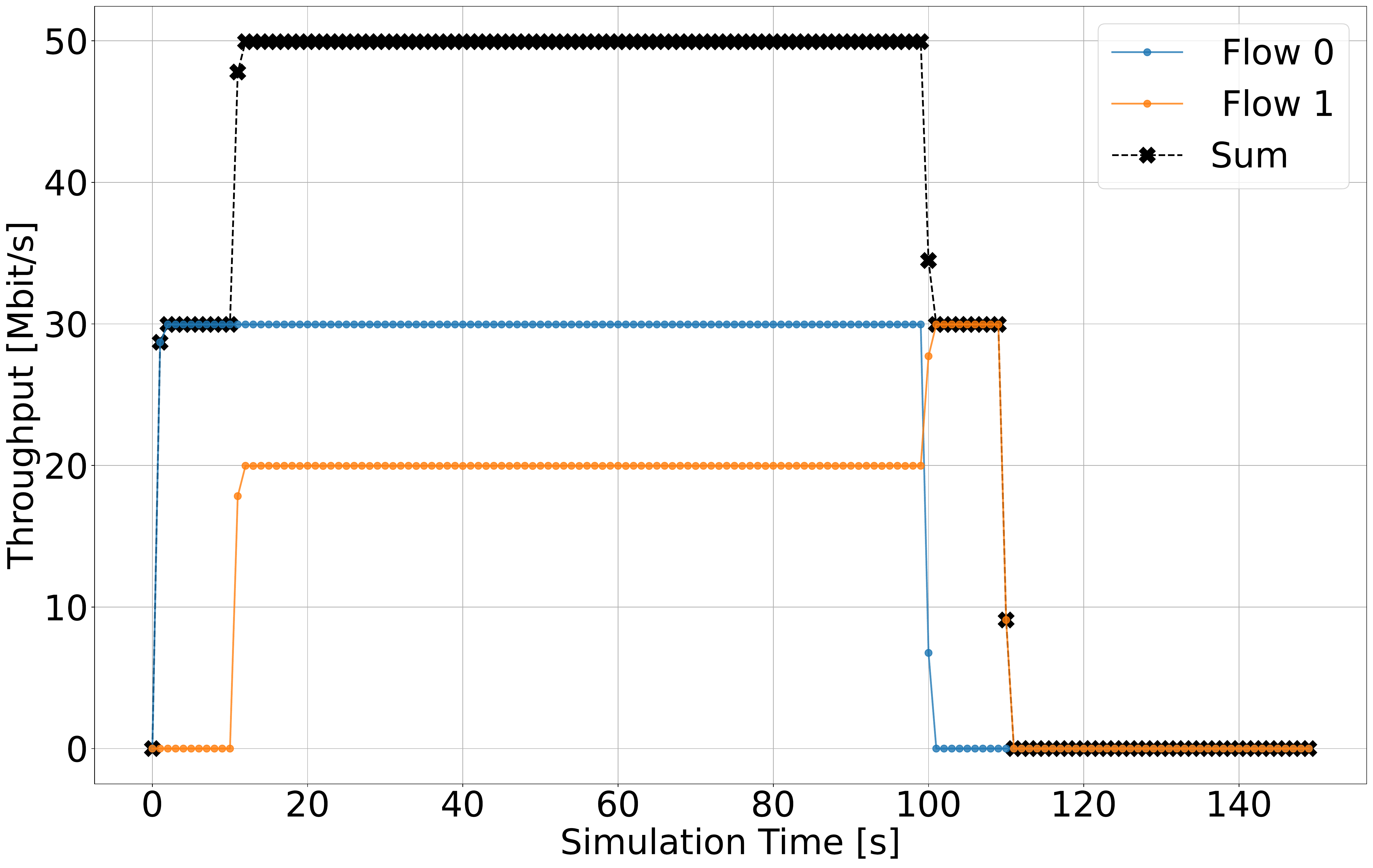}
    \caption{Scenario 3 - 2 flows with priority for flow 0.}

    \label{fig:scenarioPrio}
\end{minipage}%
 \hfill 
\begin{minipage}{.46\textwidth}
    \centering
    \includegraphics[width=1\textwidth]{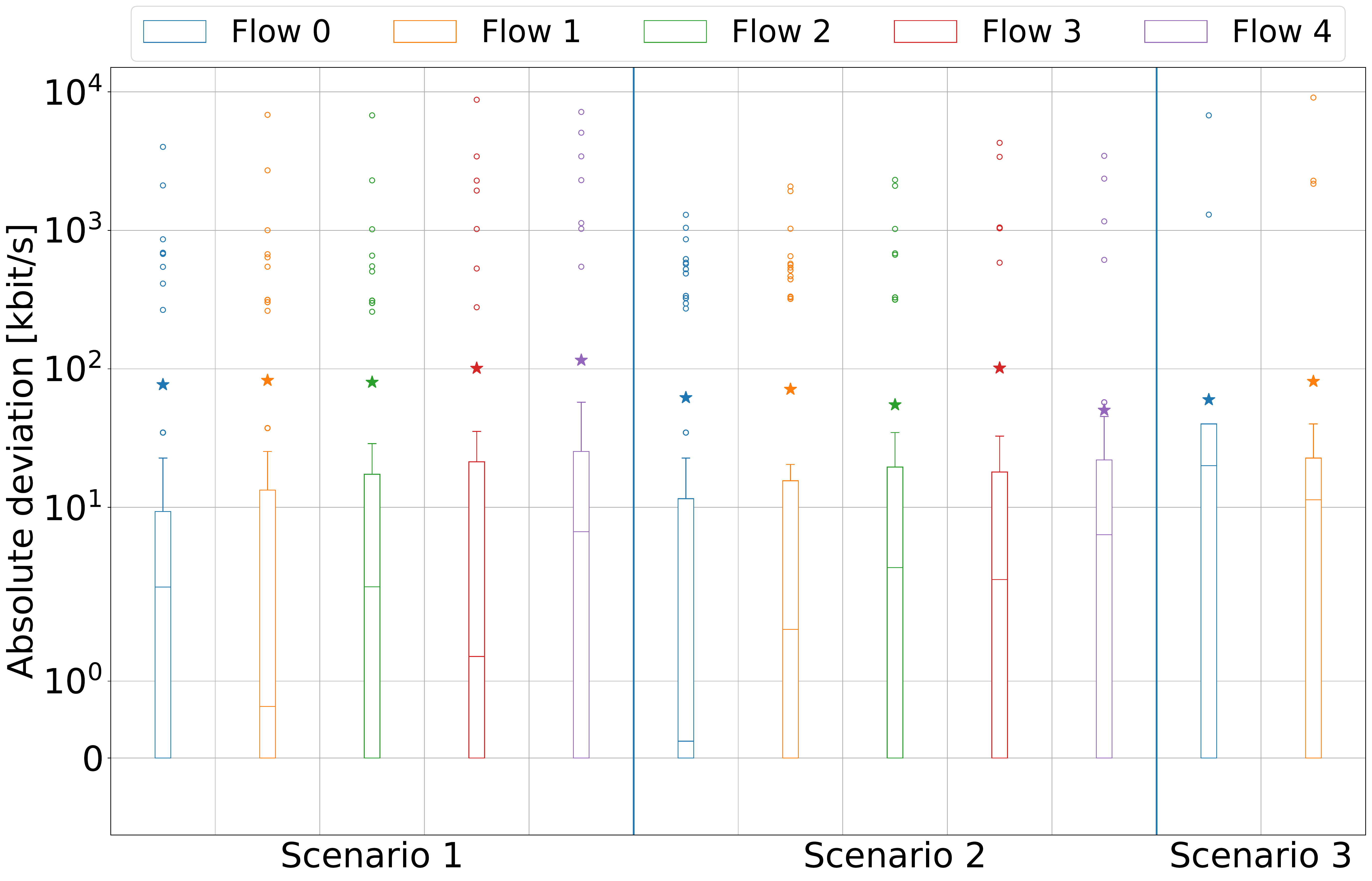}
  \caption{Absolute deviation of the obtained throughput from the expected throughput.}
  \label{fig:rate_conf}
\end{minipage}
\end{figure}



Figure~\ref{fig:scenario1} shows the results for Scenario 1. 
The plot shows the throughput of each flow on the y-axis and the x-axis represents the experiment time.
As expected, each flow receives at least its $R_{assured}$ and never exceeds its $R_{ceiling}$. 
The remaining bandwidth is also evenly divided between flows competing for it. 
In more detail, we observe that in the first 10 seconds of simulation, the only active flow 0 achieves a throughput equal to its $R_{ceiling}$, that is 20 Mbit/s. 
Flow 1, upon arrival, is able to utilize its full $R_{ceiling}$ of 25 Mbit/s as well since the total available bandwidth is not yet fully utilized. 
With the arrival of flow 2, each flow still receives its $R_{assured}$ and the remaining bandwidth is shared equally as expected. 
That is, flows 0 through 2 receive their $R_{assured}$ of 3, 6 and 9 Mbit/s, respectively and each flow additionally gets an even share of the remaining $50 - 3 - 6 - 9 = 32$ Mbit/s for a total of 13.66, 16.66, and 19.66 Mbit/s respectively.
The same principle holds until the end of the simulation.

Figure \ref{fig:scenario2} shows the results for Scenario 2. 
For the first 10 seconds, we see the same behavior as in Scenario 1. 
Between seconds 10 and 30, we see a slightly different sharing behavior, since flows 0, 1, and 2 are limited by their parent inner class.
The total bandwidth share of flows 0, 1, and 2 now cannot exceed 40 Mbit/s (the $R_{ceiling}$ of their parent).
This results in only 22 Mbit/s (instead of 32) being available for sharing and allows the flows to get 10.33, 13.33, and 16.33 Mbit/s, respectively.
Upon arrival, flow 3, having a different parent, is able to utilize 29 Mbit/s with the rates of flows 0 to 2 dropping to their share of the remaining 21 Mbit/s.
With flow 4 arriving 10 seconds later, flows 3 and 4 utilize the 30 Mbit/s ($R_{assured}$ of inner 2), each getting their $R_{assured}$ and evenly splitting the remaining bandwidth to obtain 13.5 and 16.5 Mbit/s, respectively.
Similarly, flows 0 to 2 fully utilize and share the remaining 20 Mbit/s ($R_{assured}$ of inner 1).  
Similarly, the sharing continues until the end of the experiment.

Figure \ref{fig:scenarioPrio} shows the results for Scenario 3. 
The prioritized flow 0 is able to continuously obtain its $R_{ceiling}$ and flow 1 obtains the remaining rate until flow 0 is switched off. 
The second flow is then also able to obtain its $R_{ceiling}$.
This behavior of flow 0 is in line with the expectation, as the HTB will always first satisfy the $R_{assured}$ of leaves and then satisfy the $R_{ceiling}$ of higher priority leaves.
Additionally, the end-to-end delay of flow 0 remains constant throughout the experiment at 220 ms. 
Flow 1 starts with an end-to-end delay of 320 ms and subsequently drops to 220 ms when flow 0 ends its transmission (the delays are not shown on the figures for the sake of clarity). 
This is different from the other scenarios, where we see the delay changing at every arrival or departure of a flow. Hence, priorities in the HTB can help to control the delay.  
Equivalent tests have been run using a TCP-based client and the corresponding results confirmed that the bandwidth is allocated as expected for TCP as well.



Several studies, as indicated in Section~\ref{sec:relatedwork}, highlight the high rate conformance of the HTB as one of its key properties. Therefore, we also evaluated the rate conformance obtained in the previously described scenarios. Figure~\ref{fig:rate_conf} shows the absolute deviation of the per-second throughput from the theoretically expected value, for each of the flows. For better visibility, the deviation is presented with a logarithmically scaled y-axis. A star symbol represents the mean of the absolute deviation for the corresponding flow. The horizontal lines represent the median, the boxes denote the 25th (Q1) to 75th (Q3) percentile, and the whiskers denote Q1 and Q3 plus the 1.5-fold of the inter-quartile range (Q3-Q1). For example, for Scenario 1, the mean deviations for the five flows are (76.89, 82.56, 80.03, 100.99, 115.45) kbps, respectively. Relative to the assured rates of the flows, the average deviations are equal to $2.6\%, 1.38\%, 0.89\%, 0.84\% $, and $0.77\%$, respectively. However, there are several outliers, as we can see from the figure. These outliers occur in the transient phases, i.e., upon flows arriving or leaving the system. Furthermore, HTB guarantees the rates \textit{on average}, so reaching a steady state takes longer than the 1s scale on which the measurements were done.
The median deviations are below 10~kbps for all scenarios, except for those including priorities. 
The average deviation is around 100~kbps on maximum.
Therefore, we can conclude that our \textit{HTBQueue} implementation has a high rate conformance, comparable to existing HTB implementations. 



\section{Conclusion}
\label{sec:conclusion}

The hierarchical design of HTB allows to assign each traffic flow two bandwidth parameters: an assured rate and a ceiling rate. While the first one is the minimum guaranteed rate, the second one is an upper limit until which the flow can borrow excess bandwidth. 
So far, OMNeT++ provided no possibility to configure such bitrate guarantees on a per-flow or per-class level. 
To fill this gap, we implemented the full HTB functionality as an OMNeT++ compound module, based on the implementation in the Linux traffic control (tc).
It allows to classify flows and to hierarchically configure two-level bitrate guarantees on a per-class and per-flow granularity by making use of token and ctoken buckets.
Our experimental validation shows the high conformance of assured and ceiling rates, as well as the fairness in sharing excess bandwidth between competing flows. We furthermore show that different priority levels allow the \textit{HTBQueue} to interfere in 
the fair excess bandwidth sharing mechanism, in favor of the higher priority flow. 

\section*{Acknowledgment}
The authors want to thank Martin Devera, the author of the Linux HTB, for his continuous support. 
This work is funded by the German BMBF Software Campus Grant ``BigQoE'' (01IS17052) and was supported by EC H2020 TeraFlow (101015857).
\bibliographystyle{IEEEtran}
\bibliography{refs}

\end{document}